\documentclass[prb, preprintnumbers, amsmath, amssymb, superscriptaddress, twocolumn]{revtex4}
\usepackage{graphicx, color, dcolumn, bm, natbib, amssymb, amsmath}
\usepackage[final=true]{hyperref}

\newcommand{\m}{\mathbf}

\newcommand{\p}{\parallel}
\newcommand{\ve}{\varepsilon}
\renewcommand{\d}{\dagger}

\renewcommand{\o}{\omega}

\newcommand{\be}{\begin{eqnarray}}
\newcommand{\ee}{\end{eqnarray}}
\newcommand{\nn}{\nonumber}

\begin{document}

\title{Magnons in the fan phase of anisotropic frustrated antiferromagnets}

\author{Oleg I. Utesov}
\affiliation{Department of Physics, St. Petersburg State University, St. Petersburg 198504, Russia}
\affiliation{Petersburg Nuclear Physics Institute NRC ``Kurchatov Institute'', Gatchina 188300, Russia}
\affiliation{St. Petersburg School of Physics, Mathematics, and Computer Science, HSE University, St. Petersburg 190008, Russia}

\begin{abstract}

Elementary excitations spectrum of the fan phase of anisotropic frustrated antiferromagnets is discussed analytically. In the linear spin-wave approximation, the spectrum is determined by the Hamiltonian, including normal, anomalous, and umklapp terms. The latter mix states with momenta which differ by two modulation vectors of the fan structure. This mixing leads to essential rearrangement of the low-energy part of the spectrum, which is shown to consist of a gapless phason branch with linear dispersion and a gapped ``optical'' branch, which corresponds to the fan structure amplitude oscillations. In the high-energy part of the spectrum, the effect of the umklapps is negligible, and the excitations are similar to the magnons of the fully polarized phase.

\end{abstract}

\maketitle

\section{Introduction}

Noncollinear modulated spin structures attract significant attention at the present time for various reasons. For example, such magnetic orderings can induce electric polarization in the so-called multiferroics of spin origin~\cite{KNB2005,Mostovoy,Khomskii,Arima2007,nagaosa}. Noteworthy, giant magnetoelectric effect~\cite{dzyaloshinskii1960,astrov1960} could be observed in the corresponding compounds~\cite{Kimura2003,Kenzelmann2005}, which is important for practical applications (see, e.g., review article~\cite{nagaosa} and references therein). Another point of interest is related to the topological effects. Both noncentrosymmetric compounds with Dzyaloshinskii-Moriya interaction and centrosymmetric frustrated compounds can host topologically-nontrivial magnetic structures, e.g., isolated skyrmions and skyrmion lattices~\cite{bogdanov1989,bogdanov1994,muhlbauer2009,okubo2012,leonov2015,lin2016,kurumaji2019SkL,khanh2020,bogdanov2020}. In this context, various promising applications, for instance, racetrack memory, are discussed~\cite{fert2013,fert2017}. It is also pertinent to mention the recent idea of helical structures usage as nanometer-sized inductors~\cite{nagaosa2019,kitaori2023}.

Helimagnets usually have rather complex phase diagrams on the temperature--magnetic field plane. Even in compounds that cannot host structures with several modulation vectors (such as skyrmion lattices), one can observe complicated series of phase transitions when temperature or magnetic field varies. In this context, we would like to mention multiferroics MnWO$_4$~\cite{mnw1,Ehrenberg1997,mnw2,mnw3,nojiri2011,matityahu2012,Urcelay2014} and MnI$_2$~\cite{sato,mni3,mni1,Utesov2017,UtesovMn} as particular examples. In the former, at low temperatures one can observe five magnetic field--induced phase transitions including commensurate and incommensurate spin orderings when the magnetic field is applied along the easy axis~\cite{Ehrenberg1997,nojiri2011,mitamura2012}. The corresponding theoretical description based on the anisotropic next-nearest neighbors Heisenberg (ANNNH) model was proposed in Refs.~\cite{zh,utesov2019,utesov2021phase} Importantly, in both papers incommensurate fan phase was shown to be a presaturation one (see also Ref.~\cite{Ueda2009} for the quantum isotropic model). It is pertinent to note that the fan structures can be also observed in skyrmion hosts with high-symmetry lattices~\cite{hirschberger2020,khanh2022}.

In our previous paper~\cite{utesov2021phase}, we propose an analytical approach for the fan phase static properties, including the saturation field and the field of Ising-type transition to a distorted conical helix state. In the present study, we continue our discussion and perform analytical derivation of this structure's elementary excitations spectrum. Using bosonic Holstein-Primakoff representation of the spin operators~\cite{Holstein1940} in a suitable local coordinate frame, we obtain the corresponding Hamiltonian. Noteworthy, the problem is nontrivial even at the level of the linear spin-wave approximation, since the bilinear part of the Hamiltonian consists of normal, anomalous, and umklapp terms. The latter mix excitations with momenta which differ by two spin structure modulation vectors. This mixing plays a crucial role in the low-energy part of the spectrum. It results in the emergence of gapless ``acoustic'' and gapped ``optical'' branches, which at the minima correspond to phason and standing ``breathing'' (amplitude oscillation) excitations, respectively~\cite{lee1993}. We also discuss the properties of excitation branches near the critical fields of transitions to distorted conical and fully polarized phases.

The rest of the paper is organized as follows. In Subsec.~\ref{SModel} we introduce the model under consideration. Subsec.~\ref{SGround} contains a summary of our previous results for the fan phase description. We proceed with the magnon spectrum calculations for the biaxial anisotropy in Subsec.~\ref{SMagnon}. The physical meaning of the obtained modes is discussed in Subsec.~\ref{SPhys}. We show how to adapt the developed approach for the momentum-dependent anisotropy case in Subsec.~\ref{SDip} and for other magnetic field directions in Subsec.~\ref{SMagnet}. Our findings are illustrated for a particular parameter set in Subsec.~\ref{SParam}. Finally, Sec.~\ref{SConc} contains a resume of the present study.

\section{Theory}

\subsection{Model}
\label{SModel}

We begin with a simple model of a frustrated anisotropic antiferromagnet with a small dispersionless biaxial anisotropy (adaptation of the theory for the case of momentum-dependent dipolar forces or anisotropic exchange will be discussed in Subsec.~\ref{SDip}). The models' Hamiltonian includes frustrated exchange interaction $\mathcal{H}_{\textrm{EX}}$, the anisotropy $\mathcal{H}_{\textrm{AN}}$, and the Zeeman term $\mathcal{H}_\textrm{Z}$:
\begin{eqnarray}
 \label{ham1}
  \mathcal{H} &=& \mathcal{H}_\textrm{EX} + \mathcal{H}_\textrm{AN} + \mathcal{H}_\textrm{Z}, \nonumber \\
  \mathcal{H}_\textrm{EX} &=& -\frac12 \sum_{i,j} J_{ij} \left(\mathbf{S}_i \cdot \mathbf{S}_j\right), \\
  \mathcal{H}_\textrm{AN} &=& - \sum_i \left[ D(S_i^z)^2 + E (S_i^y)^2\right], \nonumber \\
  \mathcal{H}_\textrm{Z} &=& - \sum_i \left(\mathbf{h} \cdot \mathbf{S}_i\right).\nonumber
\end{eqnarray}
Here we assume one magnetic ion per unit cell and choose $D>E>0$ for definiteness, so $z$ axis is the easy one and $x$ axis is the hard one. Symbolically we can write $D, E \ll J$, where $J$ is some characteristic energy of the exchange coupling.

The Fourier transform (we measure distances in lattice parameters, so $\m{q}$ is dimensionless)
\begin{equation}
  \label{four1}
  \mathbf{S}_j = \frac{1}{\sqrt{N}} \sum_\mathbf{q} \mathbf{S}_\mathbf{q} e^{i \mathbf{q} \mathbf{R}_j}
\end{equation}
allows us to rewrite Eqs.~\eqref{ham1} as follows:
\begin{eqnarray}
  \label{ex2}
  \mathcal{H}_\textrm{EX} &=& -\frac12 \sum_\mathbf{q} J_\mathbf{q} \left(\mathbf{S}_\mathbf{q} \cdot \mathbf{S}_{-\mathbf{q}}\right), \\
	\label{an21}
  \mathcal{H}_\textrm{AN} &=& - \sum_\mathbf{q}\left[ D S^z_\mathbf{q} S^z_{-\mathbf{q}} + E S^y_\mathbf{q} S^y_{-\mathbf{q}}\right], \\
	\label{z21}
 \mathcal{H}_\textrm{Z} &=& - \sqrt{N} \left(\mathbf{h} \cdot \mathbf{S}_{\bf 0}\right).
\end{eqnarray}
Now we can specify what kind of frustrated exchange coupling is considered. We assume that $J_\m{q}$ has two incommensurate maxima at $\m{q} = \pm \m{k}_0$ (low-symmetry lattice, for instance, orthorhombic). By virtue of this condition, we will not take into account so-called multiple-Q structures, e.g., skyrmion lattice or pre-saturation double-Q vortical phase, possible in the high-symmetry cases (see, e.g., Refs.~\cite{okubo2012,hayami2021square,utesov2021tetragonal,wang2021}). However, even in the latter case, when the external field is applied not along the high-symmetry axis, it can favor particular modulation vector~\cite{hirschberger2020,khanh2022}, and the proposed approach should be applicable too.

\begin{figure}
  \centering
  \includegraphics[width=6cm]{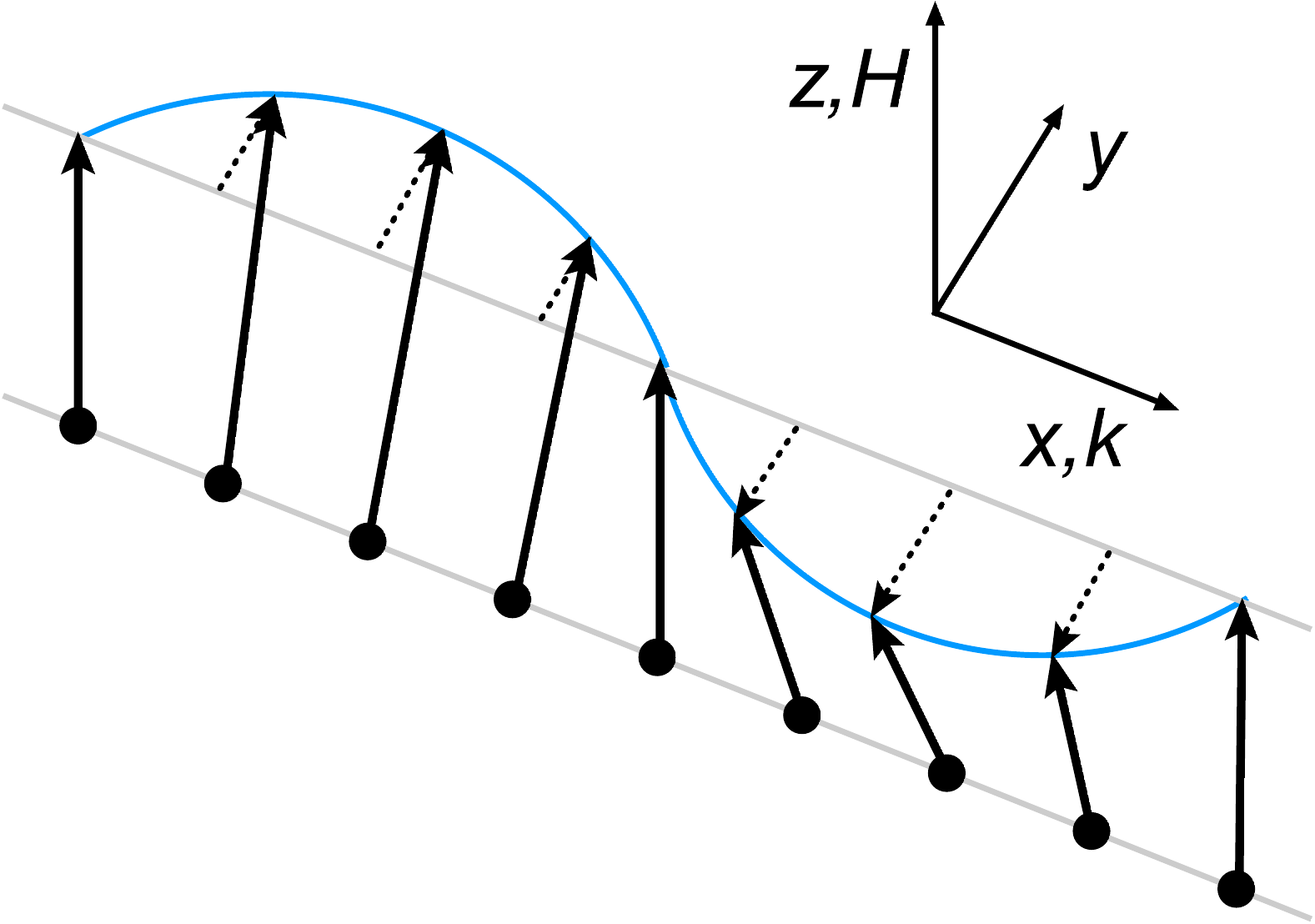}\\
  \caption{Sketch of the fan structure discussed in the present study. For definiteness and illustration purposes, magnetic field $\m{H}$ is taken along the $z$ axis, and the modulation vector $\m{k}$ is along the $x$ axis. Spins are shown by solid black arrows, their projections on the $y$ axis by dashed arrows. The blue line is the sinusoidal envelope for the latter. }\label{fig1}
\end{figure}

\subsection{Fan phase. Ground state}
\label{SGround}

Here we briefly remind the results for the fan phase ground state as well as some pertinent calculations detail from Ref.~\cite{utesov2021phase}.

Without the anisotropy, in the external field, the conical helicoid transforms into the fully polarized structure at the saturation field $h = S (J_{\m{k}_0} - J_\m{0})$. However, when the plane where spins rotate is anisotropic, the conical phase becomes unstable at a certain field, and the pre-saturation fan phase emerges, see Fig.~\ref{fig1}.

Let's, for definiteness, consider $\m{h} \p \hat{z}$. For the fan structure description, we use the following ansatz:
\begin{eqnarray}
\label{FGS1}
  \langle S^y_j \rangle &=& \beta S \cos{\mathbf{k} \mathbf{R}_j}, \nonumber\\
  \langle S^z_j \rangle &=& S \sqrt{1 -\beta^2 \cos^2{\mathbf{k} \mathbf{R}_j} }   \\
  &=& S \left[ \gamma - \varkappa \cos{2\mathbf{k}\mathbf{R}_j} - \frac{\beta^4}{64} \cos{4\mathbf{k}\mathbf{R}_j} + O(\beta^6) \right], \nonumber
\end{eqnarray}
where the small parameter $\beta \ll 1$ (which is zero in the saturated phase) is introduced and
\begin{eqnarray}
  \gamma &=& 1 - \frac{\beta^2}{4} - \frac{3 \beta^4}{64}, \\
  \varkappa &=& \frac{\beta^2}{4}\left(1 + \frac{\beta^2}{4} \right).
\end{eqnarray}
Note that in general $\m{k}$ differ with $\m{k}_0$. This issue is addressed in detail below.

According to Eqs.~\eqref{FGS1}, the local coordinate frame axes can be chosen as follows:
\begin{eqnarray} \label{Fanbasis}
  \hat{\zeta}_j &=& \hat{y} \beta \cos{\mathbf{k} \mathbf{R}_j} + \hat{z} \left[ \gamma - \varkappa \cos{2\mathbf{k}\mathbf{R}_j} - \frac{\beta^4}{64} \cos{4\mathbf{k}\mathbf{R}_j} \right], \nonumber \\
  \hat{\eta}_j &=& -\hat{y} \left[ \gamma - \varkappa \cos{2\mathbf{k}\mathbf{R}_j} - \frac{\beta^4}{64} \cos{4\mathbf{k}\mathbf{R}_j} \right] + \hat{z} \beta \cos{\mathbf{k} \mathbf{R}_j}, \nonumber\\
  \hat{\xi}_j &=& \hat{x}.
\end{eqnarray}
Here we omit $O(\beta^6)$ terms.

In order to check that our guess~\eqref{FGS1} for the fun structure ground state is correct, we should show that it results in a stable Hamiltonian. For this sake, we introduce the Holstein-Primakoff~\cite{Holstein1940} spin operators representation via bosonic ones in the following approximate form:
\begin{eqnarray}
\label{spinrep1}
  S^\zeta_j &=& S - a^\dagger_j a_j, \nonumber \\
  S^\eta_j &\simeq& \sqrt{\frac{S}{2}} \left( a_j + a^\dagger_j \right), \\
  S^\xi_j &\simeq& i \sqrt{\frac{S}{2}} \left( a^\dagger_j - a_j \right),\nonumber
\end{eqnarray}
or (in the reciprocal space)
\begin{eqnarray}
\label{spinrep2}
  S^\zeta_\m{q} &=& S \sqrt{N} \delta_{\m{q},0} - \frac{1}{\sqrt{N}} \sum_{\m{q}_1}a^\dagger_{\m{q}_1} a_{\m{q}_1+\m{q}}, \nonumber \\
  S^\eta_\m{q} &\simeq& \sqrt{\frac{S}{2}} \left( a_\m{q} + a^\dagger_{-\m{q}} \right), \\
  S^\xi_\m{q} &\simeq& i \sqrt{\frac{S}{2}} \left( a^\dagger_{-\m{q}} - a_\m{q} \right).\nonumber
\end{eqnarray}
Importantly, this approximation also allows us to discuss the magnon spectrum in the linear spin-wave theory. Next, for the spin components in the reciprocal space, one can write
\begin{eqnarray} \label{Fspin1}
  S^x_\mathbf{q} &=& S^\xi_\mathbf{q}, \nonumber \\
  S^y_\mathbf{q} &=& \frac{\beta}{2} (S^\zeta_{\mathbf{q}-\mathbf{k}} + S^\zeta_{\mathbf{q} + \mathbf{k}} ) - \gamma S^\eta_\mathbf{q}  + \frac{\varkappa}{2} (S^\eta_{\mathbf{q}-2\mathbf{k}}  \nonumber \\
  &&+ S^\eta_{\mathbf{q} + 2\mathbf{k}}) + \frac{\beta^4}{128}(S^\eta_{\mathbf{q}-4\mathbf{k}} + S^\eta_{\mathbf{q} + 4 \mathbf{k}}),  \\
  S^z_\mathbf{q} &=& \frac{\beta}{2} (S^\eta_{\mathbf{q}-\mathbf{k}} + S^\eta_{\mathbf{q} + \mathbf{k}} ) + \gamma S^\zeta_\mathbf{q} \nonumber \\
  &&- \frac{\varkappa}{2} (S^\zeta_{\mathbf{q}-2\mathbf{k}} + S^\zeta_{\mathbf{q} + 2\mathbf{k}}) - \frac{\beta^4}{128}(S^\zeta_{\mathbf{q}-4\mathbf{k}} + S^\zeta_{\mathbf{q} + 4 \mathbf{k}}). \nonumber
\end{eqnarray}
$O(\beta^6)$ terms were also omitted here.

Plugging Eqs.~\eqref{Fspin1} with the use of Eqs.~\eqref{spinrep2} into the Hamiltonian [see Eqs.~\eqref{ex2},~\eqref{an21}, and~\eqref{z21}], we obtain it in the following approximate form:
\begin{equation}
\label{Fham1}
  \mathcal{H} \approx N \epsilon_0 + \mathcal{H}_1 + \mathcal{H}_2.
\end{equation}
One sees that we do not take into account interaction terms. The classical energy per spin reads
\begin{eqnarray}
\label{Fen1}
  \epsilon_0 &=& - h S - \frac{S^2}{2}(J_\mathbf{0} + 2 D) \nonumber\\
	&&+ S \frac{\beta^2}{4}
  [h - S (J_\mathbf{k} - J_\mathbf{0} +2 E - 2 D) ] \nonumber\\
  && + S\frac{\beta^4}{64} [3 h + S(J_\mathbf{0} - J_{2\mathbf{k}})],
\end{eqnarray}
and in $\mathcal{H}_1$ we have linear terms including $a_{\pm (2n +1)\m{k}}$ and $a^\d_{\pm (2n +1)\m{k}}$ operators with coefficients $\propto \beta^{2n+1}$  ($n \geq 0$ is an integer). Analysis shows that proper choice of $\beta$ parameter which minimizes the energy~\eqref{Fen1}
\be
\label{Fbeta1}
  \beta^2 &=& \frac{8(h_s-h)}{S  (3 J_\mathbf{k} - 2 J_\mathbf{0} - J_{2\mathbf{k}})}, \, h \leq h_s, \\
  \label{hsfan}
  h_s &=& S \left( J_\m{k} - J_\m{0} + 2 E - 2 D \right)
\ee
also makes linear terms with $a_{\pm \m{k}}$ and $a^\d_{\pm \m{k}}$ vanish. Here we introduce the saturation field $h_s$ at which the fan collapses.

Importantly, other linear terms are negligible since they can alter the energy only at $O(\beta^6)$ level. Finally, we obtain the fan structure classical energy per one spin in the following form:
\begin{equation}\label{Fen2}
  \epsilon(h) = - h S - \frac{S^2}{2}(J_\mathbf{0} + 2 D) - \frac{(h_s-h)^2}{3 J_\mathbf{k} - 2 J_\mathbf{0} - J_{2\mathbf{k}}}.
\end{equation}

However, the standard ``shift in operators'' technique yields, e.g., $\langle a_{\pm 3 \m{k}} \rangle \sim \beta^3$. So, we are limited to $\beta^2$ accuracy when discussing the bilinear part of the Hamiltonian. Corresponding analysis shows that it is stable when
\begin{equation}\label{Fbcr}
  \beta^2 \leq \beta^2_{c} = \frac{8 E}{J_\mathbf{k} - J_{2 \mathbf{k}}},
\end{equation}
or equivalently  $h \geq h_{c}$, where
\begin{eqnarray}
\label{FHcr1}
  h_{c} &=& h_s - S E \frac{3J_\mathbf{k} - 2 J_ \mathbf{0}- J_{2 \mathbf{k}}}{J_\mathbf{k} - J_{2 \mathbf{k}}}.
\end{eqnarray}
is the field of transition to the distorted conical phase. Importantly, this means that in the relevant field range $\beta^2 \sim E/J \ll 1$. So, our $\beta \ll 1$ approximation relies on the anisotropy smallness assumed from the very beginning, which justifies our results above.

Here it is pertinent to discuss the fan structure modulation vector choice. Evidently, with the accuracy of our calculations at the given field $h$, we should maximize the last term in Eq.~\eqref{Fen2} for the energy. Note that both its numerator and denominator are $\m{k}$-dependent. However, since we consider fields close to the saturation one, the numerator is small, and it mostly determines the modulation vector. Particular analysis shows that under general assumptions on $J_\m{q}$ (maximum at $\m{k}_0$ and nonzero gradient of $ J_{2\mathbf{q}}$ at the same momentum) one obtains
\be \label{dk}
  \m{k} = \m{k}_0 + \delta \m{k}, \quad |\delta \m{k}| \sim \frac{h_s-h}{J} \sim \beta^2.
\ee
So, the correction is small. This property will be used below in the discussion of the magnon spectrum. Here we would like to mention that it is safe to use $\m{k}_0$ in Eqs.~\eqref{Fen2},~\eqref{Fbcr} and~\eqref{FHcr1}. Moreover, the transition field between the fan structure and the fully polarized phase is given by Eq.~\eqref{hsfan} at $\m{k} = \m{k}_0$.

\subsection{Fan phase. Elementary excitations}
\label{SMagnon}

Magnon spectrum in the linear spin-wave approximation can be derived from the bilinear part of the Hamiltonian. The latter includes normal, anomalous, and umklapp terms. Noteworthy, only the first ones contribute to the spectrum in zeroth order in E and $\beta^2$. Explicitly,
\be
  \mathcal{H}^{(0)}_2 = \sum_\m{q} S \left( J_\m{k} - J_\m{q} \right) a^\d_\m{q} a_\m{q}.
\ee
This equation is sufficient to describe the high-energy part of the spectrum: the magnon energy $\ve_\m{q} = S \left( J_\m{k} - J_\m{q} \right)$ provided that $J_\m{k} - J_\m{q} \gg E$. The latter condition can be also reformulated as ``$\m{q}$ should not be too close to $\m{k}$''.

In other case, for momenta near $\m{k}$, one should consider the Hamiltonian having the form:
\be
 \mathcal{H}_2 &=& \sum_\m{q} \Biggl[ C_\m{q} a^\d_\m{q} a_\m{q} + B_\m{q} \frac{a_\m{q} a_{-\m{q}} + a^\d_\m{q} a^\d_{-\m{q}} }{2}   \nn  \\ &&+ U_\m{q} \left(a^\d_\m{q} a_{\m{q}-2\m{k}} + a^\d_{\m{q}-2\m{k}} a_\m{q} \right) \\
 &&+ V_\m{q} \Bigl( a_\m{q} a_{-\m{q} + 2\m{k}} + a_{-\m{q}} a_{\m{q} - 2\m{k}} \nn \\ &&+ a^\d_\m{q} a^\d_{-\m{q} + 2\m{k}} + a^\d_{-\m{q}} a^\d_{\m{q} - 2\m{k}}   \Bigr) \Biggr], \nn
\ee
where
\be \label{CBUV1} \nn
C_\m{q} &=& S \Biggl[ J_\m{k} - J_\m{q} +E  \\
&&+\frac{\beta^2}{4} \left( J_\m{q} - \frac{J_{\m{q}-\m{k}}+J_{\m{q}+\m{k}}}{2} -\frac{J_{\m{k}}-J_{2\m{k}}}{2} \right) \Biggr], \nn \\
  B_\m{q} &=& S\left[ \frac{\beta^2}{8} \left( 2 J_\m{q} - J_{\m{q}-\m{k}}- J_{\m{q}+\m{k}} \right) - E \right], \\
  U_\m{q} &=& S \frac{\beta^2}{8} \left[ J_\m{k} - J_{2\m{k}} - J_{\m{q}-\m{k}} + \frac{J_{\m{q}}+J_{\m{q}-2\m{k}}}{2} \right], \nn \\
  V_\m{q} &=&   S \frac{\beta^2}{8} \frac{J_{\m{q}}-J_{\m{q}-\m{k}}}{2}. \nn
\ee
Evidently, now we have a much more complicated problem. Its exact solution hardly can be found. However, below, we show that the low-energy part of the spectrum can be obtained in an asymptotically correct fashion. Our treatment is somewhat similar to the nearly free electron model for the electron bands in solids (see, e.g., Ref.~\cite{kittel}), where the umklapps are due to the interaction with the self-consistent periodic electric potential. The latter leads to strong hybridization of the electronic states near the Brillouin zone boundary. So, in our case, the umklapps lead to substantial changes in the low-energy part of the spectrum; for the high-energy part of the spectrum, their effect is negligible being $\sim \beta^4$, except for the points where condition $|J_\m{q} - J_{\m{q} \pm 2\m{k}}| \lesssim E$ is accidentally satisfied. In the latter case, the hybridization is strong, but the corrections to the bare spectrum are still rather small $\sim \beta^2$. Below we concentrate on the low-energy part of the spectrum where some universal features can be discussed.

To perform the corresponding magnon spectrum derivation, we consider $\m{q}$ such that $J_{\m{k}_0} - J_\m{q} \lesssim E$. So $\m{q}$ is close enough to either $\m{k}_0$ or $-\m{k}_0$. Exchange interaction near its maximum can be expanded as follows:
\be \label{Jexp}
  J_{\m{k}_0 + \delta \m{q}} \approx J_{\m{k}_0} - \sum_{i=1,2,3} A_i \delta q^2_i,
\ee
where the latter tensor is written in the main axes basis. In general case, $A \sim J$ and
\be
  J_{\m{k}_0} - J_\m{q} \sim E \Longleftrightarrow |\delta \m{q}| \sim \sqrt{\frac{E}{J}}\sim \beta.
\ee
Under this assumption, we can simplify the calculations and consider only states with momenta $\m{k} + \delta \m{q}$ and $-\m{k} + \delta \m{q}$ mixing due to the umklapps. The influence of the other modes (with, e.g., $\m{q} = 3 \m{k} + \delta \m{q}$) is negligible because they belong to the high-energy sector. Note that [see Eqs.~\eqref{dk} and~\eqref{Jexp}]
\be
  J_{\m{k}+\delta\m{q}} = J_{\m{k}_0 + \delta \m{k} + \delta \m{q}} = J_{\m{k}_0+\delta\m{q}} + O(\beta^3).
\ee
By virtue of this condition, we can rewrite the coefficients of Eqs.~\eqref{CBUV1} for the two states of interest (bearing in mind their $\beta^2$ accuracy) as follows:
\be \label{CBUV2}
  \nn
  C_{\delta \m{q}} &=& S \Biggl[ \sum_{i=1,2,3} A_i \delta q^2_i + E  + \frac{\beta^2}{8} \left( J_{\m{k}} - J_\m{0}\right) \Biggr], \nn \\
  B &=& S\left[ \frac{\beta^2}{8} \left( 2 J_{\m{k}} - J_{\m{0}}- J_{2\m{k}} \right) - E \right], \\
  U &=& B + SE, \nn \\
  V &=&   S \frac{\beta^2}{16} \left( J_{\m{k}} - J_\m{0}\right). \nn
\ee

Next, we write down the Heisenberg equation of motion for the ``wave-function'' $\psi_{\delta \m{q}} = (a_{\m{k} + \delta \m{q}}, a^\d_{-\m{k} - \delta \m{q}},a_{-\m{k} + \delta \m{q}}, a^\d_{\m{k} - \delta \m{q}} )^T$, which has the following form:
\be \label{matr1}
  i \frac{\partial \psi}{\partial t} &=& \hat{M} \psi, \nn \\
  \hat{M} &=& \left(
              \begin{array}{cccc}
                C_{\delta \m{q}} & B & U & 2 V \\
                -B & - C_{\delta \m{q}} & -2 V & - U \\
                U & 2 V & C_{\delta \m{q}} & B \\
                -2 V & -U & -B & - C_{\delta \m{q}} \\
              \end{array}
            \right).
\ee
Eigennumbers of this matrix define the spectrum of the system; in order to find elementary excitations creation-annihilation operators $b^{a(o)}_{\delta \m{q}}$ and $b^{a(o)\d}_{\delta \m{q}}$ we should obtain eigenvectors $v$ of $\hat{M}^T$ and make combinations of the old bosonic operators $v^T \psi_{\delta \m{q}}$ which exponentially depend on time ($\propto e^{-i\o t}$).

The simplifications described above allow us to represent the low-energy part of the spectrum in a rather simple form:
\be \label{specA}
  \ve^{a}_{\delta \m{q}} &=& S \sqrt{A \delta q^2 \left[2E - \frac{\beta^2}{4} \left( J_{\m{k}} - J_{2 \m{k}}\right) \right]}, \\ \label{specO}
  \ve^{o}_{\delta \m{q}} &=& S \sqrt{ \left[ A \delta q^2 + 2 E \right] \left[ A \delta q^2 + \frac{2(h_s-h)}{S} \right]}.
\ee
Here we use a shorthand notation
\be
  \sum_{i=1,2,3} A_i \delta q^2_i \Leftrightarrow A \delta q^2
\ee
and introduce two branches of the spectrum: gapless ``acoustic'' $\ve^{a}_{\delta \m{q}}$ and gapped at $h<h_s$ ``optical'' $\ve^{o}_{\delta \m{q}}$. Note that the speed of magnons for the former becomes zero at $h = h_c$ [see Eq.~\eqref{Fbcr}], and the latter softens at $h=h_s$.

Importantly, one should omit $A \delta q^2$ if $\delta q \sim \beta^2$ in Eqs.~\eqref{specA} and~\eqref{specO} because we cannot control $\beta^4$ terms with the accuracy of our approach. It leads to dispersionless spectra, which looks unphysical. This issue can be solved the following way: we note that the coefficients~\eqref{CBUV1} at $\m{q} = \m{k}$ yield exactly $\ve^{a}_{\m{0}}$ and $\ve^{o}_{ \m{0}}$ given by Eqs.~\eqref{specA} and~\eqref{specO} at $\delta \m{q} = \m{0}$, respectively. So, at very small $\delta q \sim \beta^2$ the spectra given by Eqs.~\eqref{specA} and~\eqref{specO} provide reasonable interpolation between the results obtained in an asymptotically correct fashion.

\subsection{Excitations physical meaning}
\label{SPhys}

Here we discuss the question, which can be formulated as follows: ``What kind of processes do the obtained low-energy excitations represent in real space''.

First, we notice that instead of two states with $ \pm \m{k} + \delta \m{q}$, now we have acoustic and optical excitations, which can be labeled by $\delta \m{q}$ index. Note also that at $\delta \m{q} = \m{0}$ matrix $\hat{M}$ is singular [see Eqs.~\eqref{CBUV2} and~\eqref{matr1}]. Thus, we cannot construct bosonic operators for acoustic mode at this momentum and should take the $\delta \m{q} \longrightarrow 0$ limit at the final step.

Next, we take into consideration two operators:
\be
  b^a_{\delta \m{q}} = v^T\Bigl|_{\o = \ve^{a}_{\delta \m{q}} } \psi_{\delta \m{q}} \quad \text{and} \quad
  b^o_{\delta \m{q}} = v^T\Bigl|_{\o = \ve^{o}_{\delta \m{q}} } \psi_{\delta \m{q}}.
\ee
However, when treating elementary excitation operators as c-numbers, these two equations are not sufficient to determine the average values of $a$-operators. Another two equations should be written for $b^{a\d}_{-\delta \m{q}}$ and $b^{o\d}_{-\delta \m{q}}$ which contain the same set of $a$-operators.

\begin{figure}
  \centering
  \includegraphics[width=6cm]{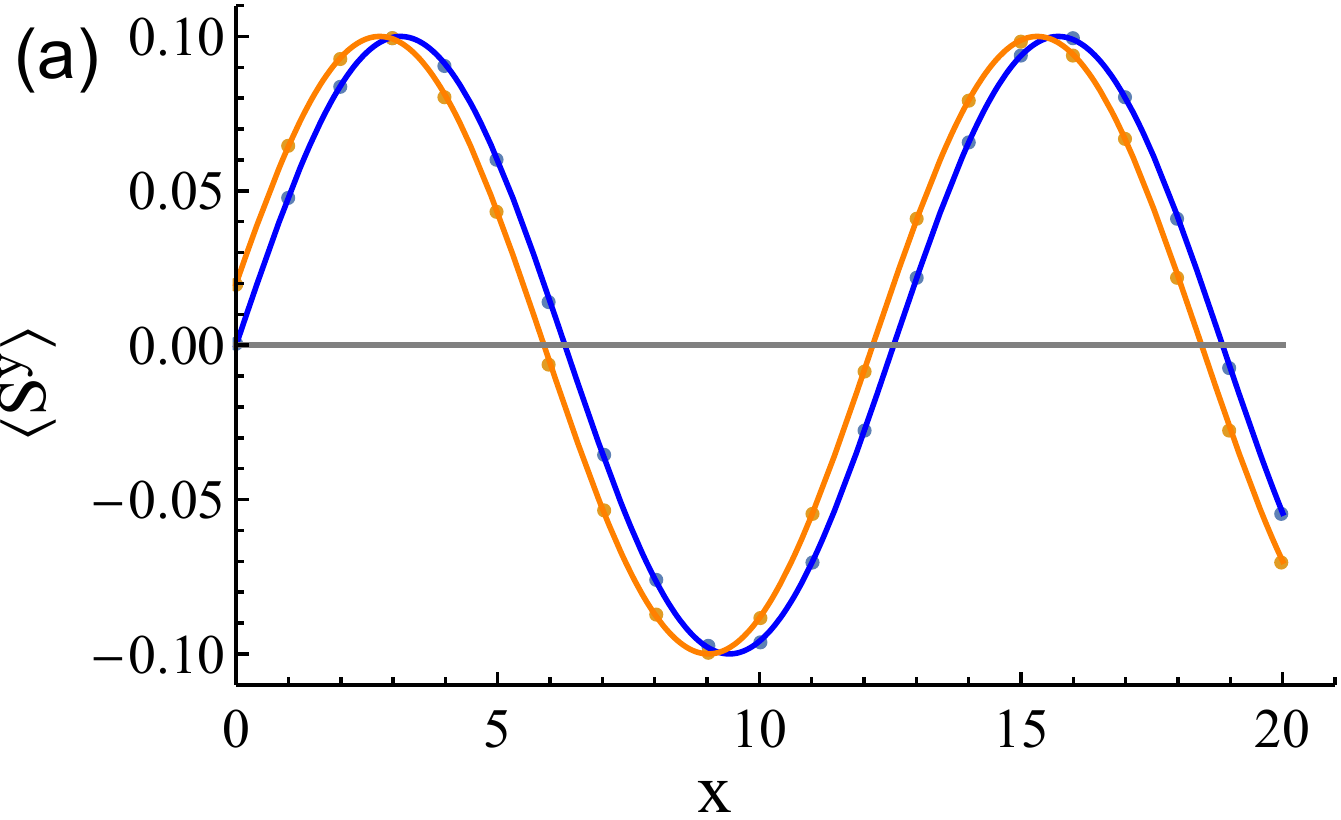} \vfill \includegraphics[width=6cm]{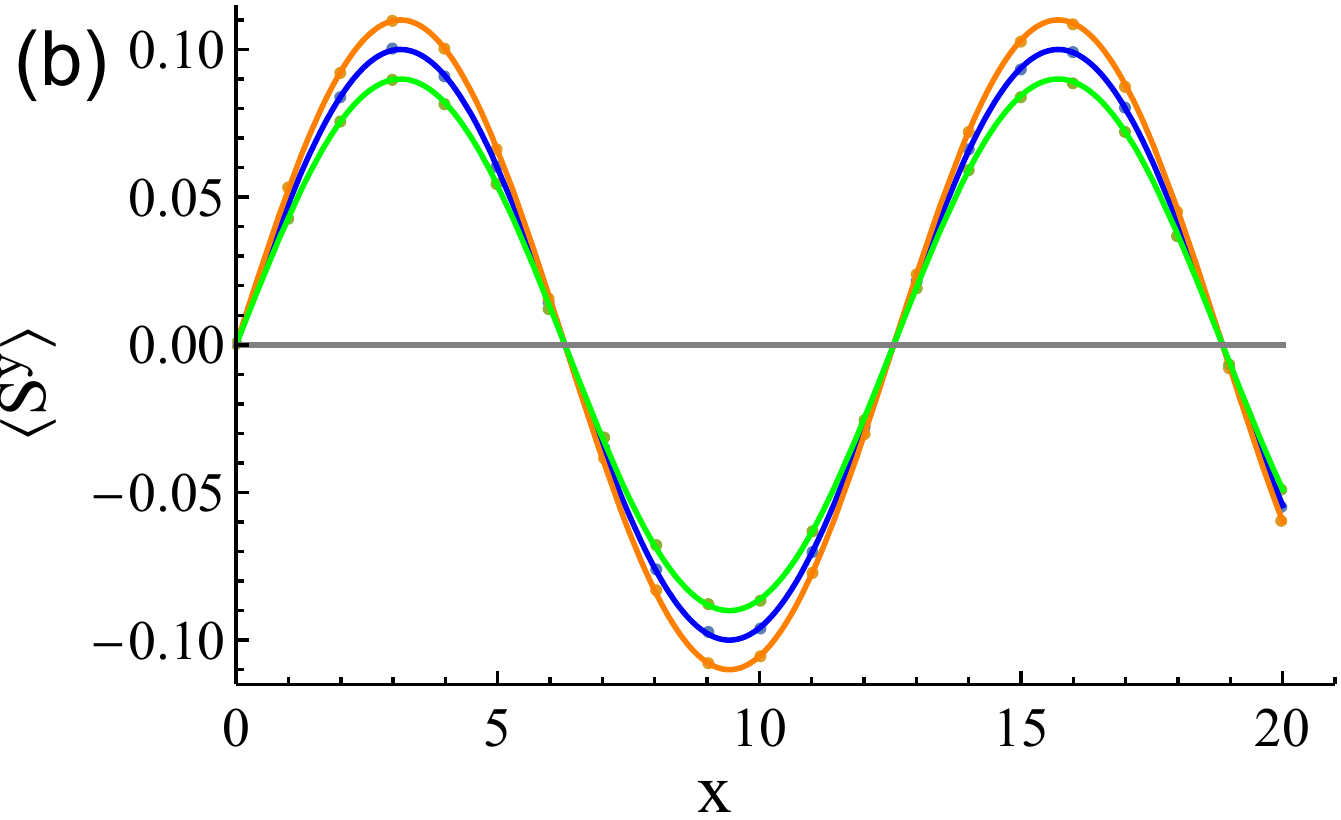} \\
  \caption{(a) Envelope functions (lines) for the spins $y$-components (dots). Phason mode at $\delta \m{q}=0$ (see text) corresponds to the homogeneous shift of the envelope function from, e.g., blue to the orange curve. (b) The same as (a), but for amplitude mode at $\delta \m{q}=0$. In this case, $\langle S^y \rangle$ harmonically oscillates between the orange and the green curves (the blue curve in the middle corresponds to the equilibrium values). For illustration purposes, the modulation vector is chosen along $\hat{x}$ in both plots. }\label{fig2}
\end{figure}

After all these preparations, we proceed with acoustic magnon at $\delta \m{q}$ and ``condense'' respective operators:
\be \label{bcond1}
  \langle b^a_{\delta \m{q}} \rangle = z,  \quad \langle b^{a\d}_{\delta \m{q}} \rangle = z^*;
\ee
all other $b$-operators average values are zero. The solution of the corresponding system of four linear equations reads
\be \label{VecPhas}
  \left(
    \begin{array}{c}
      \langle a_{\m{k} + \delta \m{q}} \rangle \\
      \langle a^\d_{-\m{k} - \delta \m{q}} \rangle \\
     \langle a_{-\m{k} + \delta \m{q}} \rangle \\
      \langle a^\d_{\m{k} - \delta \m{q}} \rangle \\
    \end{array}
  \right)  \propto z \left(
    \begin{array}{c}
      1 \\
      1 \\
      -1 \\
      -1 \\
    \end{array}
  \right).
\ee
Also, bear in mind Hermitian conjugate equality, which gives, e.g., $\langle a^\d_{\m{k} + \delta \m{q}} \rangle = z^*$. Using these values, we obtain $\langle S^\eta_j \rangle \approx \langle S^y_j \rangle \propto \textrm{Im} \, z \, \sin{\m{k}\m{R}_j}$ and $ \langle S^\xi_j  \rangle = \langle S^x_j  \rangle  = 0$, which provides a phase shift to the fan structure [cf. Eq.~\eqref{FGS1}]. So, this mode is the phason excitation~\cite{lee1993}, see Fig.~\ref{fig2}(a). Phasons play the role of Goldstone bosons of the ordered fan phase~\cite{goldstone}.

The picture above is correct until the magnetic field is stronger than $h_c$. At $h = h_c$ acoustic magnon speed becomes zero [see Eqs.~\eqref{Fbcr} and~\eqref{specA}], and small $\delta \m{q}$ expansion used above should be modified accordingly. The counterpart of Eq.~\eqref{VecPhas} is
\be \label{VecPhas2}
  \left(
    \begin{array}{c}
      \langle a_{\m{k} + \delta \m{q}} \rangle \\
      \langle a^\d_{-\m{k} - \delta \m{q}} \rangle \\
     \langle a_{-\m{k} + \delta \m{q}} \rangle \\
      \langle a^\d_{\m{k} - \delta \m{q}} \rangle \\
    \end{array}
  \right)  \propto z \left(
    \begin{array}{c}
      1 \\
      0\\
      -1 \\
      0 \\
    \end{array}
  \right).
\ee
It results in
\be
  \langle S^\eta_j \rangle &\propto&  \sin{\m{k}\m{R}_j} \, \textrm{Im} \, z , \\
  \langle S^\xi_j \rangle &\propto&  \sin{\m{k}\m{R}_j} \, \textrm{Re} \, z. \nn
\ee
Evidently, this excitation is no longer a ``pure'' phason. Condensation of such a mode in the equilibrium at $h<h_c$ leads to the formation of the distorted conical spin structure discussed in our previous paper~\cite{utesov2021phase}.

For the optical magnon, instead of the average values given by Eq.~\eqref{bcond1}, we use
\be
  \langle b^o_{\delta \m{q}} \rangle = z,  \quad \langle b^{o\d}_{\delta \m{q}} \rangle = z^*.
\ee
Repeating the same steps as for acoustic excitation, we obtain
\be
  \left(
    \begin{array}{c}
      \langle a_{\m{k} + \delta \m{q}} \rangle \\
      \langle a^\d_{-\m{k} - \delta \m{q}} \rangle \\
     \langle a_{-\m{k} + \delta \m{q}} \rangle \\
      \langle a^\d_{\m{k} - \delta \m{q}} \rangle \\
    \end{array}
  \right)  \propto z \left(
    \begin{array}{c}
      c \\
      -1 \\
      c \\
      -1 \\
    \end{array}
  \right),
\ee
where
\be
  c = \frac{B + 2 V + 2 S E + \ve^{o}_\m{0}}{B + 2V}.
\ee
Using these equations, we get
\be
  \langle S^\eta_j \rangle &\propto& (c-1) \cos{\m{k}\m{R}_j} \cos{\ve^{o}_\m{0} t}, \\
  \langle S^\xi_j \rangle &\propto& (c+1) \cos{\m{k}\m{R}_j} \sin{\ve^{o}_\m{0} t}. \nn
\ee
It can be shown that $c$ is well separated from $1$ for magnetic fields $h \in (h_c,h_s)$. So, this type of excitation can be called a ``breathing'' (or amplitude) mode since in the $yz$ plane it corresponds to the fan parameter $\beta$ oscillations [cf. Eqs.~\eqref{FGS1} and see Fig.~\ref{fig2}(b)].  Noteworthy, at $h \longrightarrow h_s$ coefficient $c \longrightarrow -1$, so $\langle S^\xi_j \rangle$ spin component vanishes when the optical magnon spectrum softens. In other cases, we observe spin precession in the $xy$ plane.

\subsection{Adaptation to momentum-dependent anisotropy}
\label{SDip}

The results above were obtained for dispersionless biaxial anisotropy. Here we show how the developed approach can be adapted for momentum-dependent anisotropy. It can be a result of magnetodipolar interaction and/or anisotropic exchange. Noteworthy, the former is always present in every real material, so let us focus on the dipolar forces. Corresponding Hamiltonian reads
\begin{equation}
 \label{Hdip1}
  \mathcal{H}_\textrm{D} = \frac12 \sum_{i,j} \mathcal{D}^{\alpha \beta}_{ij} S^\alpha_i S^\beta_j,
\end{equation}
where
\begin{equation}\label{dip1}
	 \mathcal{D}^{\alpha \beta}_{ij} = \omega_0 \frac{v_0}{4 \pi} \left( \frac{\delta_{\alpha\beta}}{R_{ij}^3} - \frac{3 R_{ij}^\alpha R_{ij}^\beta }{R_{ij}^5}\right),
\end{equation}
$v_0$ is a unit cell volume, and
\begin{equation}\label{dipen}
  \omega_0 = 4 \pi \frac{(g \mu_B)^2}{v_0} \ll J
\end{equation}
is the characteristic energy of the dipolar interaction.

After Fourier transform~\eqref{four1}, one obtains
\begin{equation}\label{dip2}
  \mathcal{H}_\textrm{D} = \frac12 \sum_\mathbf{q} \mathcal{D}^{\alpha \beta}_\mathbf{q} S^\alpha_\mathbf{q} S^\beta_{-\mathbf{q}}.
\end{equation}
At $\mathbf{q} = 0$, the tensor ${\cal D}^{\alpha \beta}_\mathbf{0}$ should be substituted by $\omega_0 \mathcal{N}_{\alpha \beta}$, where $\mathcal{N}_{\alpha \beta}$ is the demagnetization tensor~\cite{SpinWaves} and we assume the shape of the sample is ellipsoid. For each $\mathbf{q} \neq \m{0}$, symmetric tensor ${\cal D}^{\alpha \beta}_\mathbf{q}/2$ has three eigenvalues $\lambda_1(\mathbf{q}) \geq \lambda_2(\mathbf{q}) \geq \lambda_3(\mathbf{q})$ which correspond to mutually orthogonal unit vectors $\mathbf{v}_1(\mathbf{q})$, $\mathbf{v}_2(\mathbf{q})$, and $\mathbf{v}_3(\mathbf{q})$. The latter determine the hard, the middle, and the easy axes for each particular $\mathbf{q}$. Further steps are discussed below.

First, we redefine isotropic exchange as follows:
\be
  J^\prime_\m{q} = J_\m{q} - 2 \lambda_1(\m{q}).
\ee
So, in our theory instead of biaxial anisotropy~\eqref{an21}, we should include
\be
  \mathcal{H}_\textrm{AN} &=& - \sum_\m{q} \Bigl\{\left[\lambda_1(\m{q}) - \lambda_3(\m{q})\right] S^{\mathbf{v}_3}_\mathbf{q} S^{\mathbf{v}_3}_{-\mathbf{q}} \\ &&+ \left[\lambda_1(\m{q}) - \lambda_2(\m{q})\right] S^{\mathbf{v}_2}_\mathbf{q} S^{\mathbf{v}_2}_{-\mathbf{q}} \Bigr\}. \nn
\ee

Next, we note that the eigenvectors $\m{v}$ are smooth functions of $\m{q}$ in the reciprocal space and we are interested in a rather small vicinity of the fan structure modulation vector $\m{k}$ (see below). So, we can denote
\be
  \m{v}_1(\m{k}) = \hat{x}, \, \m{v}_2(\m{k}) = \hat{y}, \, \m{v}_3(\m{k}) = \hat{z}, \\
  E_\m{q} = \lambda_1(\m{q}) - \lambda_2(\m{q}), \, D_\m{q} = \lambda_1(\m{q}) - \lambda_3(\m{q}).
\ee
Evidently, $D_\m{q}>E_\m{q}>0$, and our current problem becomes very similar to the one solved above with the dispersionless anisotropy. We define $\m{k}_0$ as a momentum which maximizes the combination
\be
   J^\prime_\m{q} + 2E_\m{q} = J_\m{q} - 2 \lambda_2(\m{q})
\ee
so that, for instance, the field of the transition to the fully polarized phase reads [cf. Eq.~\eqref{hsfan}]
\be
  h_s &=& S \left( J^{\prime}_{\m{k}_0} - J^\prime_\m{0} + 2 E_{\m{k}_0} - 2 D_\m{0} \right).
\ee
Once again, actual modulation vector $\m{k}$ differs by a small $\delta \m{k}$ from $\m{k}_0$, see Eq.~\eqref{dk}. So, in various static properties discussed in Subsec.~\ref{SGround} one should use $E_\m{k}$ instead of $E$ and $D_\m{0}$ instead of $D$.

When the dynamic properties are addressed, it can be shown that in Eqs.~\eqref{CBUV1} several changes are in order:
\be \label{CBUV3}
  C_\m{q} &=& S \Biggl[ J^\prime_\m{k} - J^\prime_\m{q} + 2 E_\m{k} - E_\m{q}  \\
&&+\frac{\beta^2}{4} \left( J_\m{q} - \frac{J_{\m{q}-\m{k}}+J_{\m{q}+\m{k}}}{2} -\frac{J_{\m{k}}-J_{2\m{k}}}{2} \right) \Biggr], \nn \\
  B_\m{q} &=& S\left[ \frac{\beta^2}{8} \left( 2 J_\m{q} - J_{\m{q}-\m{k}}- J_{\m{q}+\m{k}} \right) - E_\m{q} \right],
\ee
while $U_\m{q}$ and $V_\m{q}$ stay intact (note that the change $J_\m{q} \rightarrow J^\prime_\m{q}$ in terms $\propto \beta^2$ has no sense with the accuracy of our calculations). One can see that the first line of $C_\m{q}$ equals to
\be
  (J^\prime_\m{k} + 2 E_\m{k}) - (J^\prime_\m{q} + 2 E_\m{q}) + E_\m{q}
\ee
and in the small $\delta \m{q} = \m{q} - \m{k}$ limit it can be rewritten as follows:
\be
  \sum_{i=1,2,3} A_i \delta q^2_i + E_\m{q}
\ee
Moreover, there is also no need to track $\delta \m{q}$ dependence of $E_\m{q}$, so we can just use the constant $E_{\m{k}_0}$ instead.

Finally, we see that all the required parameters for magnon spectra calculations have absolutely the same form as in the case of dispersionless anisotropy, thus the respective results are correct in this case too. It should be only pointed out that $\m{k}_0$ is defined by $J^\prime_\m{q} + 2 E_\m{q}$ maximum, and $A_i$ coefficients are determined by this combination expansion for small $\delta \m{q}$. Note that reformulation of the theory for the anisotropic exchange can be done the same way as for dipolar forces.

\subsection{Other directions of magnetic field}
\label{SMagnet}

Noteworthy, the approach developed above can be generalized for magnetic fields along other than easy directions. We discuss the necessary alterations for the dispersionless case, where the notation is simpler.

If the external field is applied along the middle $\hat{y}$ axis, one should just interchange $D$ and $E$ parameters. In this case, $D$ measures the in-plane anisotropy, and one will get different phase  transitions fields $h_s$ and $h_c$ [see Eqs.~\eqref{hsfan} and~\eqref{FHcr1}]. Next, to obtain the magnon spectrum, one should plug $D$ instead of $E$ into Eqs.~\eqref{specA} and~\eqref{specO}. Note that axes $\hat{z}$ and $\hat{y}$ should be also swapped.

At the same level of simplicity, we can consider the magnetic field along the hard $\hat{x}$ axis. To make the connection with our previous result transparent, we add a constant term
\be
  \sum_i E \left[ (S^x_i)^2 + (S^y_i)^2 + (S^z_i)^2 \right]
\ee
to the Hamiltonian~\eqref{ham1}. So, it becomes evident that one can use the results obtained above with the following substitutions:
\be
  D \rightarrow -E, \quad E \rightarrow D - E,
\ee
bearing in mind the axes redesignation
\be
  \hat{x} \rightarrow \hat{y}, \, \hat{y} \rightarrow \hat{z}, \,  \hat{z} \rightarrow \hat{x}.
\ee

\subsection{Particular parameter set}
\label{SParam}

\begin{figure}[t]
  \centering
  \includegraphics[width=6.5cm]{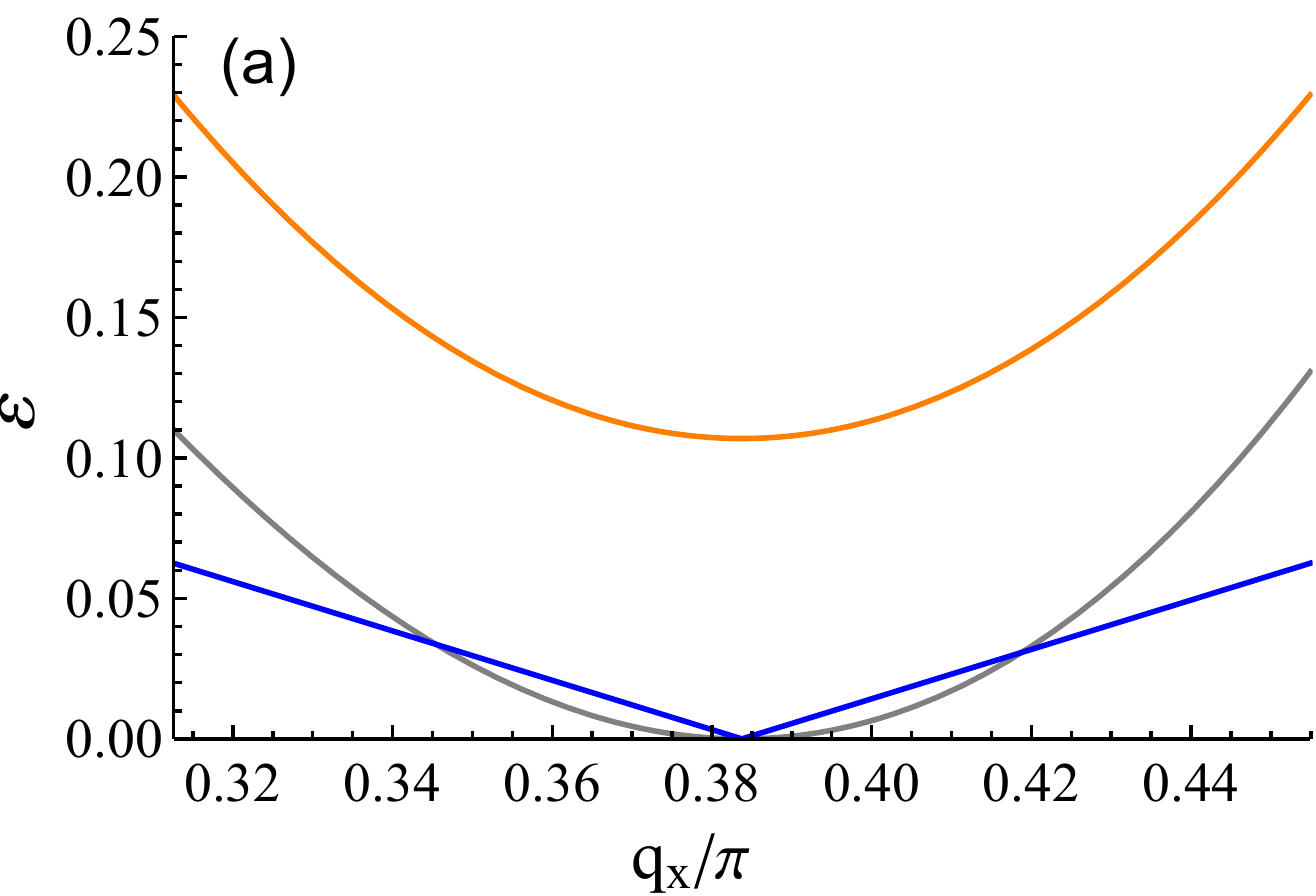} \vfill \centering \includegraphics[width=6cm]{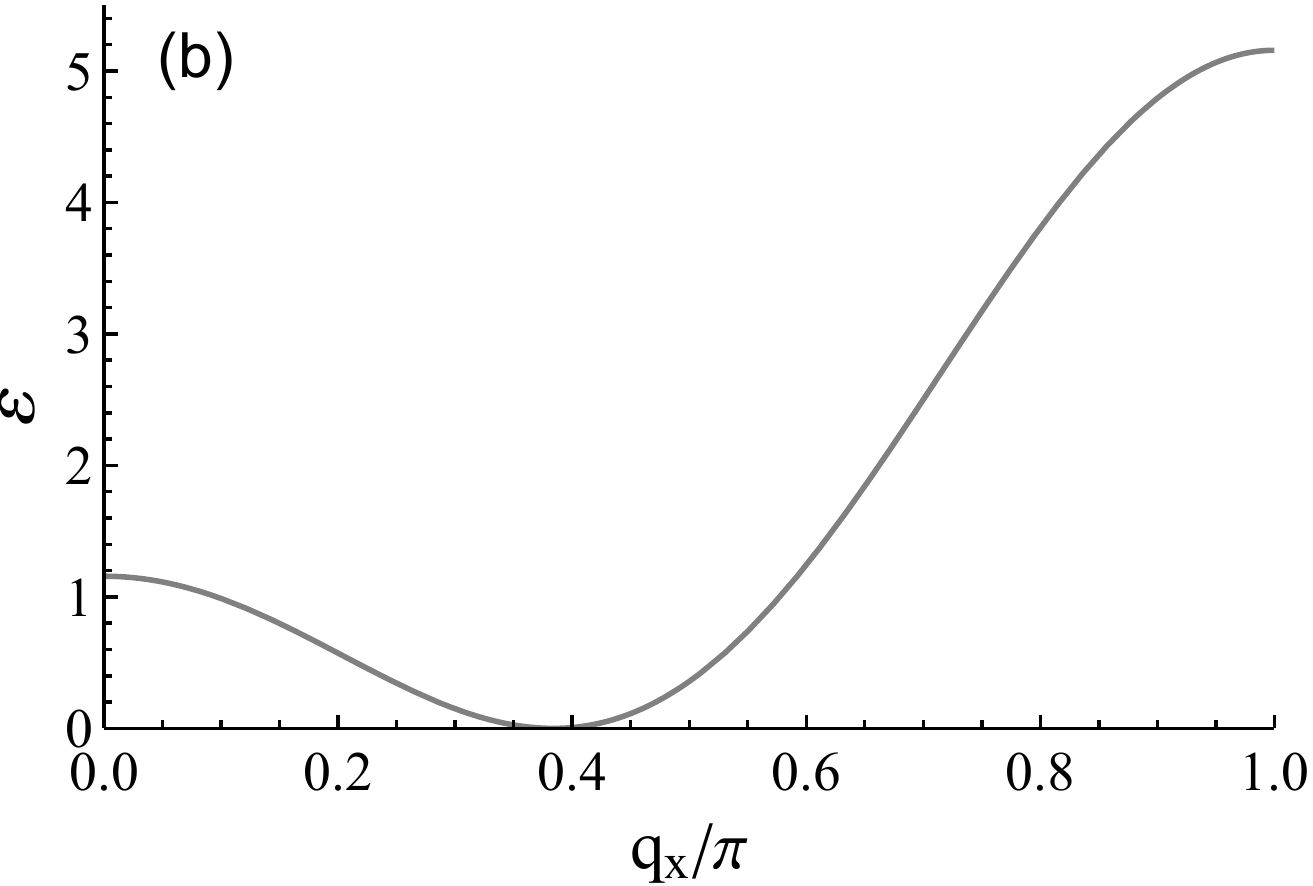} \\
  \caption{(a) Magnon spectra in the vicinity of the fan structure modulation vector $\m{k}_0$ for a particular parameter set (see text). Here the umklapp processes become crucial and lead to substantial changes in the spectrum. The blue curve stands for the acoustic branch, the orange one for the optical branch, and the gray curve represents the bare spectrum. (b) The bare magnon spectrum in the half of the first Brillouin zone. For the most part of it, the umklapps are negligible. }\label{fig3}
\end{figure}

As an example, we consider the magnon spectrum of the fan phase of the anisotropic next-nearest neighbors Heisenberg model (ANNNH).

Let the system consists of ferromagnetically coupled spin chains oriented along the $x$ axis. The exchange interaction is frustrated along the chains and can be represented in the form
\be
  J_\m{q} = J_1 \cos{q_x} + J_2 \cos{2 q_x} + J_{\textrm{FM}}(q_y,q_z),
\ee
where the last term stands for not frustrated ferromagnetic inter-chain exchange, which has a maximum at $q_y=q_z=0$. We take $J_1=1$, $J_2=-0.7$, so $\m{k}_0 \approx (0.38 \pi, 0, 0)$. We also choose the biaxial anisotropy parameters $D=0.1$ and $E=0.05$. Then, for the external magnetic field along the $z$ axis, we have $h_c \approx 0.97$, $h_s \approx 1.06$.

In Fig.~\ref{fig3}(a), we compare strongly renormalized due to the umklapps magnon spectrum, including gapless linear acoustic~\eqref{specA} and gapped optical~\eqref{specO} branches with the  ``bare'' one (calculated without corrections due to nonzero $E$ and $\beta$) in the $\sqrt{E}$ vicinity of the $\m{k}_{0}$ point obtained at $h=1$ ($\beta^2 \approx 0.08$) as functions of $q_x$ ($q_y=q_z=0$). Fig.~\ref{fig3}(b) shows the bare spectrum in the half of the first Brillouin zone. If $\ve_\m{q} \gg E$ ($q_x$ not very close to $k_{0x}$) the umklapps-induced corrections are negligible.

\section{Conclusions}
\label{SConc}

To conclude, we study the elementary excitations spectrum of the fan spin structure, which can be observed as a presaturation one in various anisotropic centrosymmetric helimagnets. We show that the Hamiltonian, sufficient to describe the magnon spectrum in the linear spin-wave theory, consists of normal, anomalous, and umklapp terms. The latter are responsible for the substantial rearrangement of the low-energy part of the spectrum. It consists of the gapless ``acoustic'' branch with linear dispersion [see Eq.~\eqref{specA}] and the ``optical'' one with the gap [see Eq.~\eqref{specO}]. The former corresponds to the so-called phason mode, which should emerge in the ordered phase in accordance with Goldstone's theorem; the latter can be called a ``breathing'' mode since it corresponds to the fan structure amplitude oscillations.

Noteworthy, at the critical field of transition to distorted conical structure, the speed of acoustic excitations becomes zero, and they no longer correspond to simple phasons. The component along the hard axis emerges in this case. So, the condensation of such excitations leads to the formation of the distorted conical spiral. In contrast, at the saturation field, the optical branch softens, and the spectrum becomes doubly degenerate.

The high energy part of the spectrum is mainly determined by the normal terms, whereas the anomalous and the umklapp contributions are negligible. In this case, the elementary excitations are similar to the magnons of the fully polarized phase, which correspond to a simple coherent precession of the spins.

\begin{acknowledgments}

This work is supported by the Russian Science Foundation (Grant No. 22-22-00028).

\end{acknowledgments}

\bibliography{TAFbib}

\end{document}